\documentclass{ws-ijmpa}
\usepackage{latexsym}
\usepackage{graphicx,epsfig}

\begin{document}
\markboth{Rakhi et al.}
{A Cosmological Model with Fermionic Field and Gauss-Bonnet Term}

\catchline{}{}{}{}{}
%%%%%%%%%
\title{A COSMOLOGICAL MODEL WITH FERMIONIC FIELD AND GAUSS-BONNET TERM}

\author{RAKHI R.\footnote{corresponding author.}}
\address{School of Pure and Applied Physics, Mahatma Gandhi University, Priyadarshini Hills Post,\\Kottayam, Kerala, India. PIN 686 560.
\\rakhir006@gmail.com}

\author{G. V. VIJAYAGOVINDAN \footnote{deceased}}
\address{School of Pure and Applied Physics, Mahatma Gandhi University, Priyadarshini Hills Post,\\Kottayam, Kerala, India. PIN 686 560.}

\author{NOBLE P.ABRAHAM}
\address{School of Pure and Applied Physics, Mahatma Gandhi University, Priyadarshini Hills Post,\\Kottayam, Kerala, India. PIN 686 560.
\\noblepa@gmail.com}

\author{INDULEKHA K.}
\address{School of Pure and Applied Physics, Mahatma Gandhi University, Priyadarshini Hills Post,\\Kottayam, Kerala, India. PIN 686 560.
\\kindulekha@gmail.com}

\maketitle

\begin{history}
\received{}
\revised{}
\end{history}

\begin{abstract}
In this work, a cosmological model inspired by string theory with Gauss-Bonnet term coupled to the fermionic field is taken into consideration. The self-interaction potential is considered as a combination of the scalar and pseudo-scalar invariants. Here the cosmological contribution of the coupling of Gauss-Bonnet term with a non-Dirac fermionic field--characterized by an \textit{interaction term }$L_{DG} ^2$--  is investigated. It is observed that the new type of coupling plays a significant role in the accelerating behavior of the universe. Specifically, in addition to the late time acceleration for the universe, $L_{DG} ^2$ produces an early decelerating behavior. The behavior of the equation-of-state parameter $\left( w \right)$ is such that it guarantees the stability of the theory.

\end{abstract}

\ccode{PACS numbers: 98.80.-k, 98.80.Cq, 98.80.Qc}

\section{Introduction}

Nojiri et.al \cite{nojiri} proposed the Gauss-Bonnet dark energy model, inspired by string/M-theory where standard gravity with scalar contains additional scalar-dependent coupling with Gauss-Bonnet invariant. Their study indicates that current acceleration may be significantly influenced by string effects. Also, the current acceleration of the Universe may be caused by a mixture of scalar phantom. Moreover, they concluded that scalar-Gauss-Bonnet coupling acts against the occurrence of Big Rip  in phantom cosmology. Ribas et.al \cite{ribas}, in 2005, investigated whether fermionic field, with a self-interacting potential that depends on scalar and pseudo-scalar invariants, could be responsible for accelerated periods during the evolution of the Universe, where a matter field
would answer for the post inflation decelerated period. They have shown that the fermionic field behaves like an inflation field for the early Universe and later on,
as a dark energy field, whereas the matter field was created by an irreversible process connected with a non-equilibrium pressure.

It is expected that the combination of fermionic field and Gauss-Bonnet term will produce significant effects in cosmological evolution. Therefore, we propose a cosmological model with fermionic field and Gauss-Bonnet term. The coupling of Gauss Bonnet term to the fermionic field is introduced in order to produce a late time acceleration for the universe. We explore whether the combined effect can produce an early inflation as well as late time acceleration. Taking inspiration from string/M-theory in which additional higher order terms are required, a term
proportional to the square of the fermionic Lagrangian density,which we call as an \textit{interaction term}, is added to the total Lagrangian. In this model, the additional coupling is suppressed by some powers of the string scale. So we produce a long term effect without disturbing the overall stability. It may be noted that the fermion condensate expected from Super Symmetry theory also calls for higher order terms.

\section{String Theory Inspired Cosmological Model with Fermionic Field and Gauss-Bonnet Term}

The action for this model reads
\begin{equation}
\label{eq1}
S=\int {\sqrt {-g} } d^4x\left( {L_g +L_m +a_1 L_{DG} +a_2 L_{DG}^2} \right)
\end{equation}
where $L_g =R/2$, with $R$ denoting the curvature scalar, $L_g$ is the gravitational Einstein Lagrangian density, $L_m $ is the Lagrangian of the matter field and $a_1 $and $a_2 $ are coupling constants. Finally, the augmented Dirac Lagrangian density $L_{DG}$ for a fermionic mass $m$ is given by
\begin{equation}
\label{eq2}
L_{DG} =\frac{i}{2}\left[ {\overline \Psi \Gamma ^\mu D_\mu \Psi -\left( {D_\mu
\overline \Psi } \right)\Gamma ^\mu \Psi } \right]-m\left\langle {\overline
\Psi \Psi } \right\rangle -V L_{GB}
\end{equation}
In Eq. (\ref{eq2}), $V=V(\Psi,\overline \Psi )$ describes the potential density of self-interaction between fermions, $L_{GB} $ is the Gauss-Bonnet Lagrangian density\cite{nojiri}, given by $L_{GB} =R^2 - 4R_{\mu \nu } R^{\mu \nu } + R_{\mu \nu \rho \sigma } R^{\mu \nu \rho \sigma }$ and $\overline \Psi =\Psi ^\dag \gamma ^0$ denotes the adjoint spinor field. Moreover, the connection between general relativity and Dirac equation is done via the tetrad formalism and the components of the tetrad play the role of gravitational degrees of freedom. That is, $\Gamma ^\mu =e_a^\mu \gamma ^a$ are the generalized Dirac-Pauli matrices, where $e_a^\mu $ denote the tetrad or \textit{vierbein}. The covariant derivatives \cite{ribas} in Eq. (\ref{eq2}) are given by,
\[
D_\mu \Psi =\partial _\mu \Psi -\Omega _\mu \Psi
\]
\begin{equation}
\label{eq3}
D_\mu \overline \Psi =\partial _\mu \overline \Psi +\overline \Psi \Omega _\mu
\end{equation}
with the spin connection,
\begin{equation}
\label{eq4}
\Omega _\mu =-\frac{1}{4}g_{\rho \sigma } [\Gamma _{\mu \delta }^\rho
-e_b^\rho \partial _\mu e_\delta ^b ]\Gamma ^\sigma \Gamma ^\delta
\end{equation}
where $\Gamma _{\mu \delta }^\rho $ is the Christoffel symbol.

\subsection{Basic Equations}
Based upon the cosmological principle, which states that our universe is homogenous and isotropic, we use the Robertson-Walker metric to describe our Universe:
\begin{equation}
\label{eq5}
ds^2=dt^2-a(t)^2\left( {dx^2+dy^2+dz^2} \right)
\end{equation}

Through Euler-Lagrange equations, from Eqs. (\ref{eq1}) and (\ref{eq2}), we can obtain the equations of motion for the spinor field as
\begin{equation}
\left(a_1+2 a_2 L_{DG}\right) dL_{DG}/d\overline \psi=0
\end{equation}which requires that either $\left(a_1+2 a_2 L_{DG}\right)=0$
  or   $dL_{DG}/d\overline \psi=0$.

The first case, i.e. $\left(a_1+2 a_2 L_{DG}\right)=0$ , gives $L_{DG}=-a_1/2 a_2$.Being forced to have $L_{DG}$ equal to a constant is an unphysical and artificial constraint. Hence we
rule out that option.

Now choosing the second case, we get the Dirac equations for the spinor field as $dL_{DG}/d\overline \psi=0$ or
\begin{equation}
\label{eq6}
i\Gamma ^\mu D_\mu \Psi -m\Psi -L_{GB} \frac{dV}{d\overline \Psi }=0
\end{equation}
The equation of motion is exactly the same as in the case where no $L_{DG}^2$ term
is present. However we see that the $L_{DG}^2$ term modifies the source terms for gravity
in a significant way. We also note that any series of arbitrary powers $L_{DG}^n$ present
in the original Lagrangian would lead to $\delta L_{DG}/\delta \psi =0$ for the fermionic equation of motion, because the other equation arising would be unphysical. So our analysis can
be thought of as treating the case of a general form for the fermionic contribution to the stress tensor, where we retain terms only upto the next to leading order contribution in an expansion of the stress tensor in the derivatives of $\psi$.

Similarly, we can obtain the Dirac equations for the adjoint coupled to the
gravitational field.

Also the variation of the action (Eq.(\ref{eq1})) with respect to the tetrad gives the
Einstein's field equations as
\begin{equation}
\label{eq7}
R_{\mu \nu } -\frac{1}{2}g_{\mu \nu } R=-T_{\mu \nu }
\end{equation}
where $T_{\mu \nu } =T_f^{\mu \nu } +T_m^{\mu \nu } $, with $T_f^{\mu \nu } $ being the energy-momentum tensor of the fermionic field and $T_m^{\mu \nu} $, that of the matter field. The symmetric form of the energy-momentum tensor of the fermionic field \cite{green,birrell} which follows from Eq. (\ref{eq1}) gives
\[
T_f^{\mu \nu } =\frac{i}{4}\left\{ {\overline \Psi \Gamma ^\mu D^\nu \Psi
+\overline \Psi \Gamma ^\nu D^\mu \Psi -D^\nu \overline \Psi \Gamma ^\mu
\Psi -D^\mu \overline \Psi \Gamma ^\nu \Psi } \right\}+\frac{i}{2}L_{DG}
\{\overline \Psi \Gamma ^\mu D^\nu \Psi +\overline \Psi \Gamma ^\nu D^\mu
\Psi -
\]
\begin{equation}
\label{eq8}
D^\nu \overline \Psi \Gamma ^\mu \Psi -D^\mu \overline \Psi \Gamma ^\nu \Psi
\}-g^{\mu \nu }L_{DG} -g^{\mu \nu }L_{DG} ^2
\quad
\end{equation}
where the Dirac matrices have the form $\Gamma ^0=\gamma ^0,\Gamma
^i=\frac{1}{a(t)}\gamma ^i,\Gamma ^5=\gamma ^5$ and the spin connection
components are
\begin{equation}
\label{eq10}
\Omega _0 =0,\Omega _i =\frac{1}{2}\mathop a\limits^. (t)\gamma ^i\gamma ^0
\end{equation}
From Eq. (\ref{eq8}), together with Eqs. (\ref{eq1}), (\ref{eq2}) and (\ref{eq5}), we can obtain the
non-vanishing components of the energy-momentum tensor for the fermionic
field as
\begin{equation}
\label{eq10ab}
\left( {T_f } \right)_0^0 =a_1 \left( {m\left\langle {\overline \Psi \Psi }
\right\rangle +VL_{GB} } \right)-a_2 \left( {m\left\langle {\overline \Psi
\Psi } \right\rangle +VL_{GB} } \right)^2+a_2 \left( {m\left\langle
{\overline \Psi \Psi } \right\rangle +\frac{\overline \Psi
}{2}\frac{dV}{d\overline \Psi }L_{GB} +\frac{\Psi }{2}\frac{dV}{d\Psi
}L_{GB} } \right)^2
\end{equation}
\begin{equation}
\label{eq10a}
(T_f )_1^1 =(T_f )_2^2 =(T_f )_3^3 =a_1 L_{GB} \left[ {V-\frac{\overline
\Psi }{2}\frac{dV}{d\overline \Psi }-\frac{dV}{d\Psi }\frac{\Psi }{2}}
\right]-a_2 L_{GB}^2 \left[ {V-\frac{\overline \Psi }{2}\frac{dV}{d\overline
\Psi }-\frac{dV}{d\Psi }\frac{\Psi }{2}} \right]^2
\end{equation}
The interaction between the fermionic constituents are modeled through the
presence of a non-equilibrium pressure term$\left( \varpi \right)$ in the
source's energy-momentum tensor. Therefore, for an isotropic, homogenous
Universe, the components of the energy-momentum tensor of the fermionic
field can be identified as 
\begin{equation}
\label{eq10b}
(T_f )_\nu ^\mu =diag\left( {\rho _f ,-p_f ,-p_f
,-p_f } \right). 
\end{equation}
One can obtain from Eqs. (\ref{eq10ab}), (\ref{eq10a}) and (\ref{eq10b}), the energy density and pressure of the fermionic part:
\begin{equation}
\label{eq11}
\rho _f =a_1 (m\left\langle {\overline \Psi \Psi } \right\rangle +VL_{GB}
)-a_2 (m\left\langle {\overline \Psi \Psi } \right\rangle +VL_{GB} )^2+a_2
(m\left\langle {\overline \Psi \Psi } \right\rangle +\frac{\overline \Psi
}{2}\frac{dV}{d\overline \Psi }L_{GB} +\frac{\Psi }{2}\frac{dV}{d\Psi
}L_{GB} )^2
\end{equation}
\begin{equation}
\label{eq12}
-p_f =a_1 L_{GB} \left[ {V-\frac{\overline \Psi }{2}\frac{dV}{d\overline
\Psi }-\frac{dV}{d\Psi }\frac{\Psi }{2}} \right]-a_2 L_{GB}^2 \left[
{V-\frac{\overline \Psi }{2}\frac{dV}{d\overline \Psi }-\frac{dV}{d\Psi
}\frac{\Psi }{2}} \right]^2
\end{equation}
The self-interaction potential takes the form,
\begin{equation}
\label{eq13}
V=\lambda \left[ {\beta _1 \left( {\left\langle {\overline \Psi \Psi }
\right\rangle } \right)^2+\beta _2 \left( {i\left\langle {\overline \psi
\gamma ^5\Psi } \right\rangle } \right)^2} \right]^n
\end{equation}
where $\lambda $ is the coupling constant and $n$ is a constant exponent. We
consider $V$ as a combination of the scalar and pseudo-scalar invariants.
That is, $\beta _1 =\beta _2 =1$. 

Combining  Eqs. (\ref{eq11}), (\ref{eq12}) and (\ref{eq13}) we solve for the energy density and
pressure of the fermionic field as
\begin{equation}
\label{eq14}
\rho _f =a_1 \left( {m\left\langle {\overline \Psi \Psi } \right\rangle
+VL_{GB} } \right)-a_2 \left( {m\left\langle {\overline \Psi \Psi }
\right\rangle +VL_{GB} } \right)^2+a_2 \left( {m\left\langle {\overline \Psi
\Psi } \right\rangle +2nVL_{GB} } \right)^2
\end{equation}
\begin{equation}
\label{eq15}
p_f =a_1 \left( {2n-1} \right)VL_{GB} +a_2 \left[ {\left( {2n-1}
\right)VL_{GB} } \right]^2
\end{equation}
respectively.
We infer from Eq.(\ref{eq15}) that the pressure is negative only for $n <1/2$ and this could represent either the inflaton or the dark energy.

The Friedman equations give
\begin{equation}
\label{eq16}
3H^2=\rho =\rho _f +\rho _m
\end{equation}
\begin{equation}
\label{eq16a}
2\frac{\mathop a\limits^{..} }{a}=-\frac{\rho }{3}-p-\varpi =-H^2-p-\varpi
\end{equation}
or 
\begin{equation}
\label{eq16b}
2\frac{\mathop a\limits^{..} }{a}+H^2=-p_f -p_m -\varpi 
\end{equation}
Using Eqs. (\ref{eq16}) and (\ref{eq16b}), the acceleration equation becomes
\begin{equation}
\label{eq17}
\frac{\mathop a\limits^{..} }{a}=\frac{2H^2-\rho _f -\rho _m -p_f -p_m
-\varpi }{2}
\end{equation}
The conservation law for the energy density of the fermionic field and
matter field can be written as
\begin{equation}
\label{eq18}
\mathop {\rho _f }\limits^. +3H\left( {\rho _f +p_f } \right)=0
\end{equation}
\begin{equation}
\label{eq19}
\mathop {\rho _m }\limits^. +3H\left( {\rho _m +p_m +\varpi } \right)=0
\end{equation}
where $H=\frac{\mathop {a(t)}\limits^. }{a}$is the Hubble's parameter.

Considering the fermionic field to be a function of time alone, the Dirac
equations (\ref{eq5}) becomes
\begin{equation}
\label{eq20}
\mathop \Psi \limits^. +\frac{3}{2}H\Psi +im\gamma ^0\Psi +i\gamma ^0L_{GB}
\frac{dV}{d\overline \Psi }=0
\end{equation}
Similarly, the Dirac equation for the adjoint spinor field coupled to the
gravitational field becomes
\begin{equation}
\label{eq21}
\mathop {\overline \Psi }\limits^. +\frac{3}{2}H\overline \Psi -im\overline
\Psi \gamma ^0-iL_{GB} \frac{dV}{d\overline \Psi }\gamma ^0=0
\end{equation}

\subsection{Field Equations}

The system of field equations that help us to find the cosmological
solutions is:

\subsubsection{The Acceleration Equation }
The acceleration equation for the cosmological model under consideration can
be obtained from equations (\ref{eq14}), (\ref{eq15}) and (\ref{eq17}). Also we consider the
pressure of the matter field as $p_m =w_m \rho _m $ with $0\le w_m \le 1$,
which is a barotropic equation of state. We again consider a vanishing
energy density of the matter field $\left( {\rho _m } \right)$ and a
vanishing non-equilibrium pressure $\left( \varpi \right)$at t=0.
Incorporating all these ideas, the acceleration equation for the present
model becomes
\begin{equation}
\label{eq22}
\frac{\mathop a\limits^{..} }{a}=c\pm d
\end{equation}
where \begin{equation}
\label{eq22a}
c=\frac{1}{2}\frac{0.007}{a_2 H^4V^2}\left( {1+7.2a_1 H^2V-9.6a_2H^2Vm\left\langle {\overline \Psi \Psi } \right\rangle } \right)
\end{equation} and 
\begin{equation}
\label{eq23}
d=\frac{1}{2}\sqrt {\frac{4.9E-5}{a_2^2 H^8V^4}\left( {1+7.2a_1 H^2V-9.6a_2
H^2Vm\left\langle {\overline \Psi \Psi } \right\rangle } \right)^2-\left[
{\frac{0.015}{a_2 H^4V^2}\left( {2H^2-a_1 m\left\langle {\overline \Psi \Psi
} \right\rangle -\rho _m -w_m \rho _m -\varpi } \right)} \right]}
\quad
\end{equation}

\subsubsection{Evolution equation for the energy density of matter field}
The evolution equation for $\rho _m $ is given by
\begin{equation}
\label{eq24}
\mathop {\rho _m }\limits^. +3H(\rho _m +w_m \rho _m +\varpi )=0
\end{equation}

\subsubsection{Evolution equation for the non-equilibrium pressure}

The linearised form of the evolution equation for non-equilibrium pressure\cite{kremer1,kremer2,kremer3} reads
\begin{equation}
\label{eq25}
\tau \mathop \varpi \limits^. +\varpi +3\eta H=0
\end{equation}

\subsubsection{The Dirac Equation }
In terms of the spinor components,$\Psi =\left( {\Psi _1 ,\Psi _2 ,\Psi _3
,\Psi _4 } \right)^T$, the Dirac Eq. (\ref{eq20}) can be written as
\begin{equation}
\label{eq26}
\frac{d}{dt}\left( {\begin{array}{l}
 \Psi _1 \\
 \Psi _2 \\
 \Psi _3 \\
 \Psi _4 \\
 \end{array}} \right)+\frac{3}{2}H\left( {\begin{array}{l}
 \Psi _1 \\
 \Psi _2 \\
 \Psi _3 \\
 \Psi _4 \\
 \end{array}} \right)+im\left( {\begin{array}{l}
 \Psi _1 \\
 \Psi _2 \\
 -\Psi _3 \\
 -\Psi _4 \\
 \end{array}} \right)-48i\frac{\mathop a\limits^{..} }{a}H^2\left( {\Psi
_1^\dag \Psi _1 +\Psi _2^\dag \Psi _2 -\Psi _3^\dag \Psi _3 -\Psi _4^\dag
\Psi _4 } \right)\left( {\begin{array}{l}
 \Psi _1 \\
 \Psi _2 \\
 -\Psi _3 \\
 -\Psi _4 \\
 \end{array}} \right)\frac{dV}{d\left( {\overline \Psi \Psi } \right)^2}
\quad
\end{equation}
$-48i\frac{\mathop a\limits^{..} }{a}H^2\left( {\Psi _3^\dag \Psi _1 +\Psi
_4^\dag \Psi _2 -\Psi _1^\dag \Psi _3 -\Psi _2^\dag \Psi _4 } \right)\left(
{\begin{array}{l}
 \Psi _3 \\
 \Psi _4 \\
 -\Psi _1 \\
 -\Psi _2 \\
 \end{array}} \right)\frac{dV}{d\left( {\overline \Psi \gamma ^5\Psi }
\right)^2}=0
$
\subsubsection{Equation -of -State Parameter }
The equation-of -state parameter $(w)$ is given by
\begin{equation}
\label{eq27}
w=\frac{p}{\rho }
\end{equation}
This should be thought of as a phenomenological relation reflecting the
current amount of pressure and energy density in the dark energy.

\subsection{Cosmological Solutions}

In order to obtain numerical solutions of the coupled system of Eqs. (\ref{eq22})-
(\ref{eq27}), we have to specify the parameters $\lambda ,\beta _1 ,\beta _2
,n,m,\alpha ,w_m $ and $a_1 $. These parameters take the following values:

$\lambda=0.1 ,\beta _1=\beta _2=1,n=0.3,m=0.01,\alpha=1,\varpi=1$ and $a_1=1$.
The initial conditions, we have chosen for t=0 (by adjusting clocks) are: $a\left( 0
\right)=1,\Psi _1 \left( 0 \right)=0.1i,\Psi _2 \left( 0 \right)=1,\Psi _3
\left( 0 \right)=0.3,\Psi _4 \left( 0 \right)=i,\rho _m \left( 0
\right)=0,\varpi \left( 0 \right)=0.$ The last two initial conditions correspond to a vanishing energy density of the matter field and a vanishing non-equilibrium pressure at t = 0. The conditions chosen here characterize qualitatively an initial proportion between the constituents in the corresponding era.

Graphs are plotted for various values of $a_{2}$.
The effect of $L_{DG} ^2$ is now studied first by choosing $a_2 =1$[Figs.1-5].
\subsubsection{Acceleration }
In Fig 1, we plot the accleration as the root of (\ref{eq22}), corresponding to the upper + sign, whereas in Fig 2, the acceleration corresponding to the lower - sign is plotted.

\begin{figure}[htbp]
\centerline{\includegraphics[width=2.53in,height=1.5in]{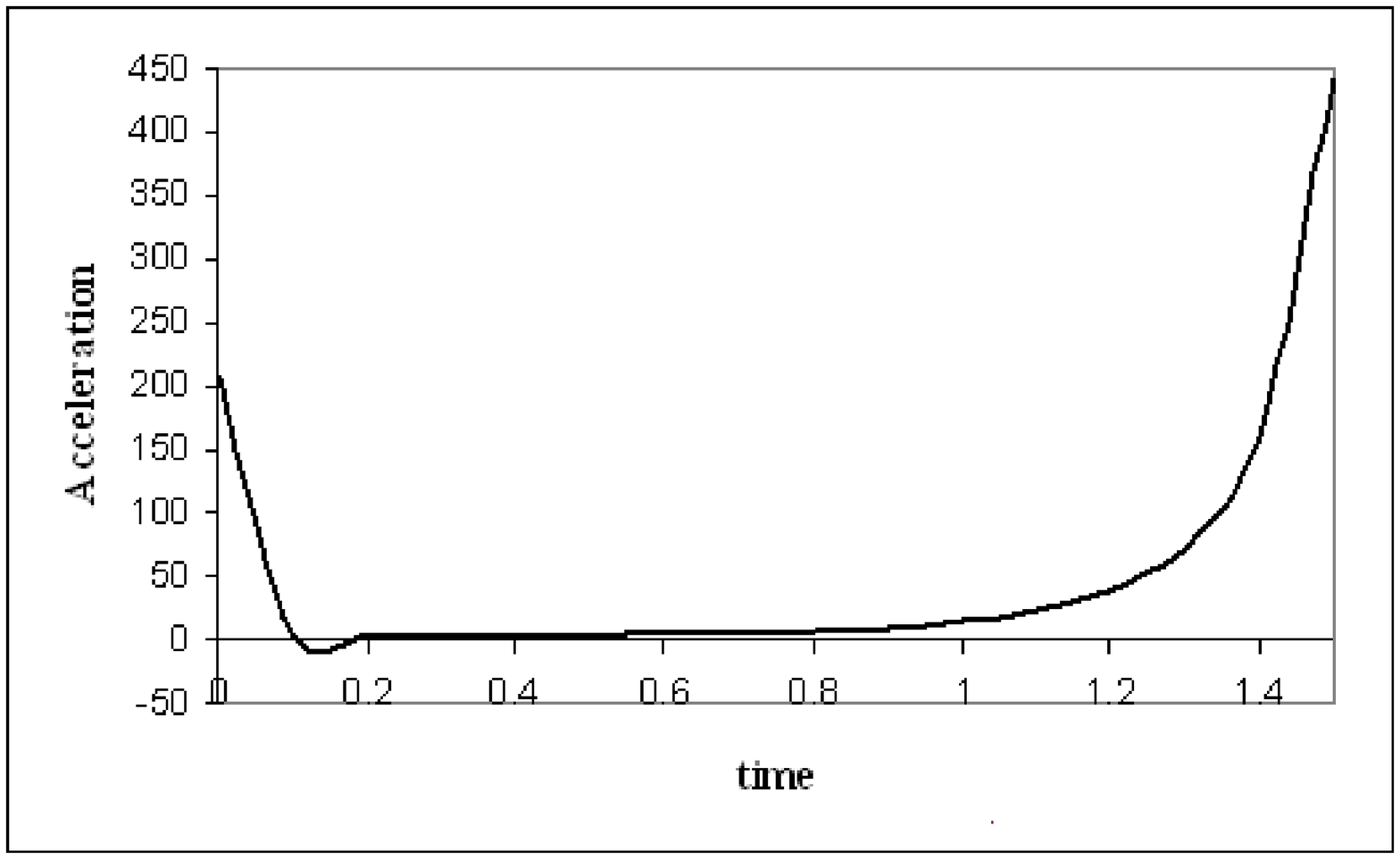}}
\label{fig1}
\caption{}
\end{figure}
\begin{figure}[htbp]\centerline{\includegraphics[width=2.53in,height=1.50in]{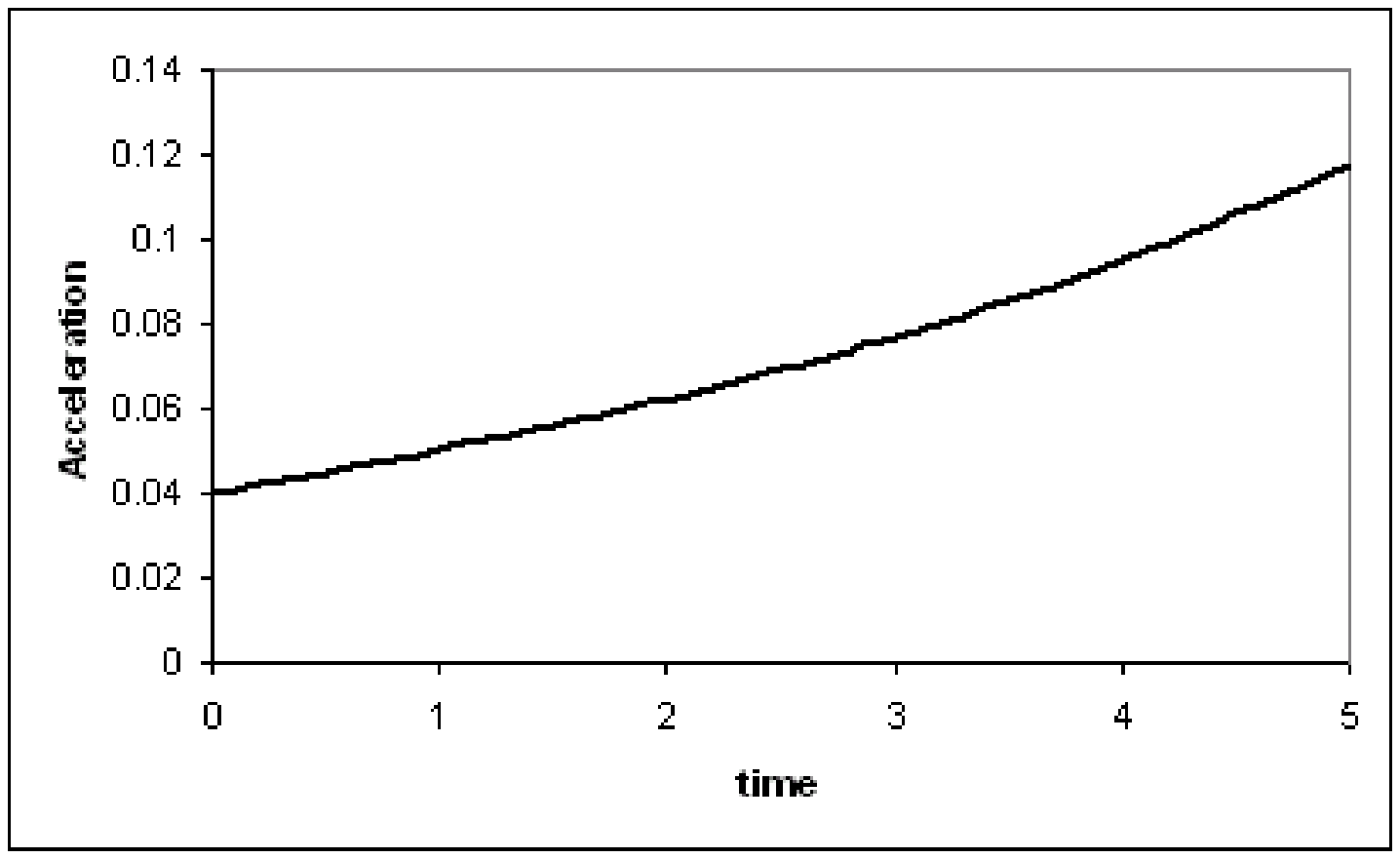}}
\label{fig2}
\caption{}
\end{figure}

From the following Fig 2, it is clear that the negative root has no interesting features. Hence from now onwards we concentrate only on the root corresponds to the + sign.

\subsubsection{Energy density}
In Figs. 3 and 4, we plot the evolution of energy densities of fermionic field and matter field respectively. 
\begin{figure}[hbtp]
\centerline{\includegraphics[width=2.53in,height=1.5in]{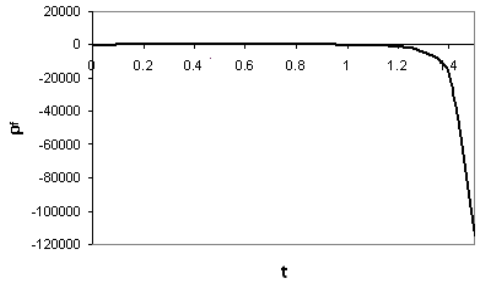}}
\label{fig3}
\caption{Energy density of fermionic field $\rho _{f}$Vs time}
\end{figure}
\begin{figure}[h]
\centerline{\includegraphics[width=2.53in,height=1.5in]{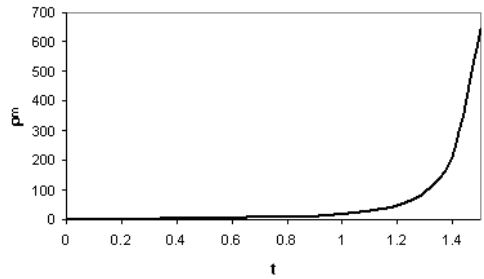}}
\label{fig4}
\caption{Energy density of matter field $\rho _{m}$Vs time}
\end{figure}
\subsubsection{Equation-of-State Parameter}
The evolution of equation-of-state parameter is plotted below:
\begin{figure}[htbp]
\centerline{\includegraphics[width=2.53in,height=1.5in]{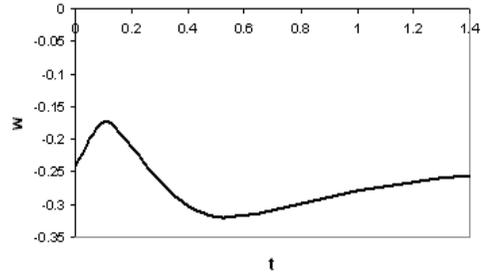}}
\label{fig5}
\caption{Equation of State Parameter of the universe Vs time}
\end{figure}

The effect of $L_{DG}^2$ is studied below, by putting $a_2 =$ 5, 10, 50,100 successively [Figs.6-9].
\begin{figure}[htbp]
\centerline{\includegraphics[width=2.83in,height=1.8in]{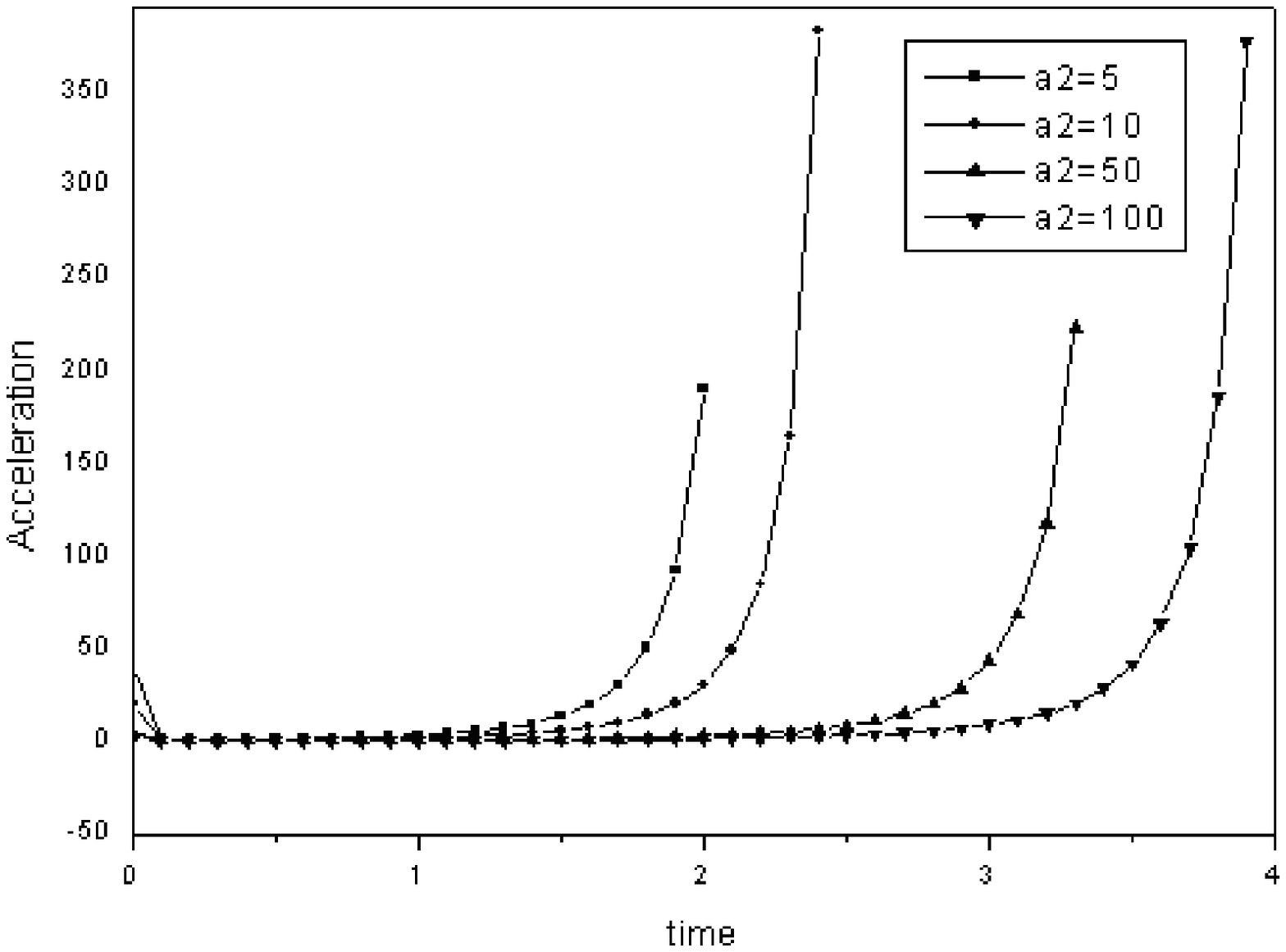}}
\label{fig6}
\caption{Acceleration Vs time for various $a_2 $}

\centerline{\includegraphics[width=2.93in,height=1.86in]{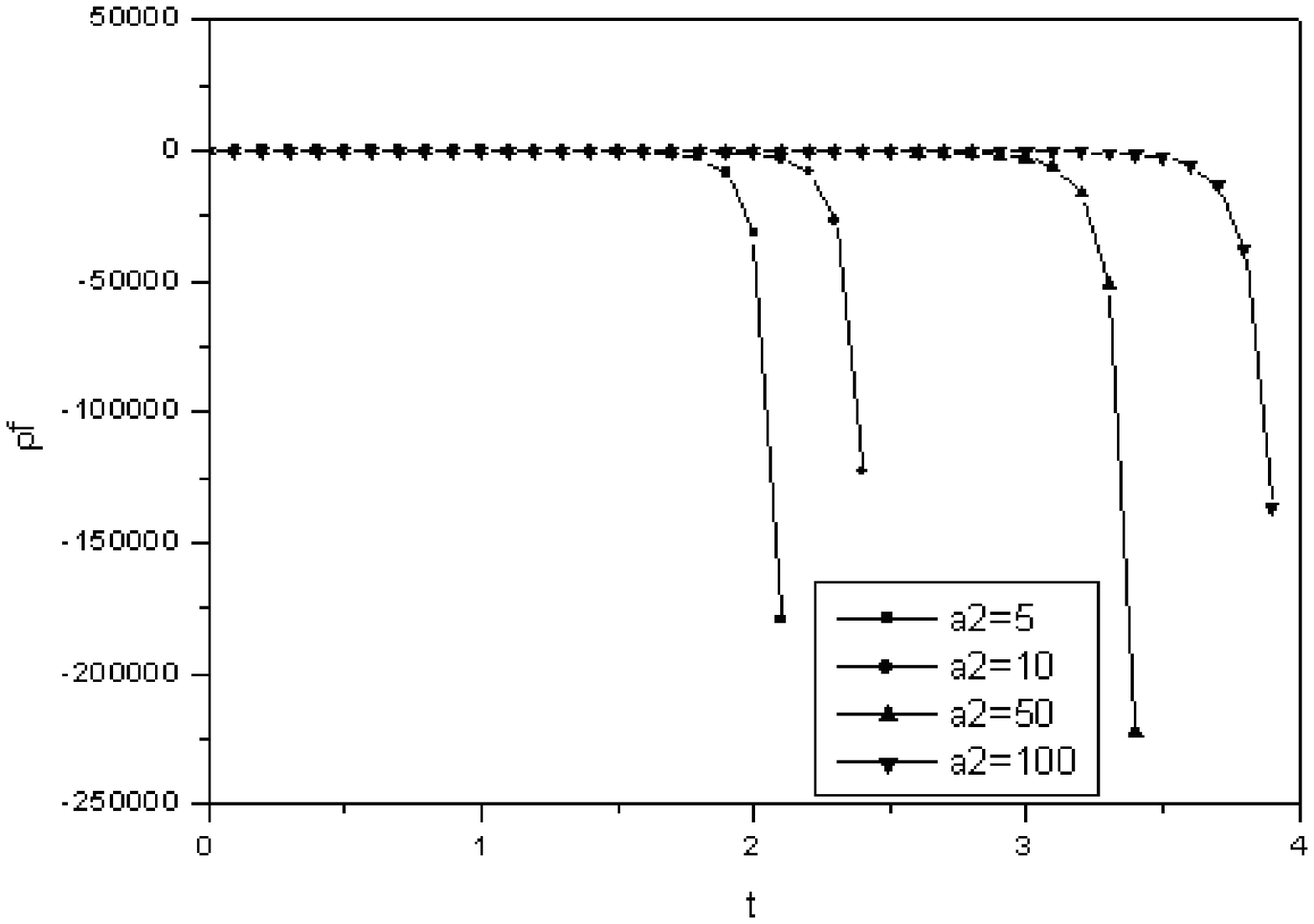}}
\label{fig7}
\caption{Energy density of the fermionic field $\rho _f $ Vs time for various $a_2 $}
\end{figure}
\begin{figure}[h]
\centerline{\includegraphics[width=2.86in,height=1.80in]{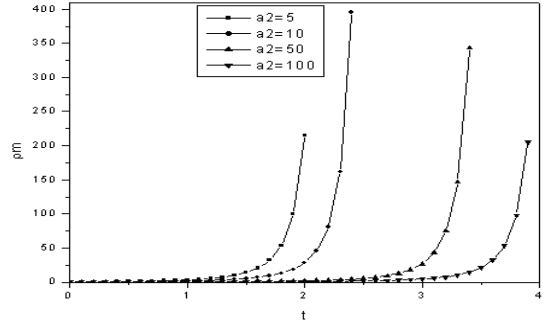}}
\label{fig8}
\caption{Energy density of the matter field $\rho _m$ Vs time for various $a_2 $}
\end{figure}
\begin{figure}[htbp]
\centerline{\includegraphics[width=2.88in,height=1.85in]{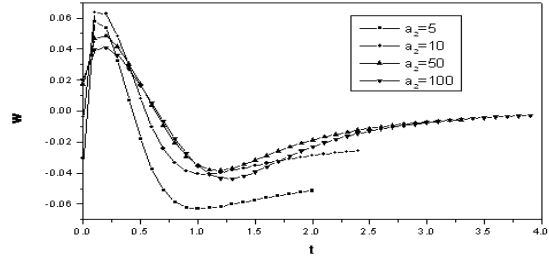}}
\label{fig9}
\caption{Equation-of-state parameter of the Universe Vs time for various $a_2 $}
\end{figure}
\subsection{Discussions}

\subsubsection{Acceleration}
From the above obtained graphs, it is seen that the coupling of Gauss-Bonnet term with non-Dirac fermionic field produces an early inflation preceded by a short term deceleration. At later times, there is an increase in the acceleration effect. This late time acceleration is considered to be the effect of Gauss-Bonnet term. Also as the value of the coefficient of $L_{DG}^2$, $a_2 $ increases, the initial acceleration decreases and hte late time acceleration becomes more delayed. This shows that $L_{DG}^2$ has a prominent role in the present model. It is possible to get the purely non- Dirac fermionic field, by simply putting $a_1 =0$.

\subsubsection{Energy density / Equation of state parameter}
It is evident from the plots that the energy density of the matter field evolves more rapidly for smaller values of a$_{2}$. As a$_{2}$
increases, the late time evolution obtained for $\rho _{m}$ becomes more and more delayed. Also it is observed that the decay of the energy density of the fermionic field comes earlier for smaller values of $a_{2}$, causing larger accelerated period. As the value of $a_{2}$ increases, the decay of $\rho _{f}$ starts at a later time.

These effects can be explained on the basis of equation-of-state parameter. The equation-of-state parameter is connected directly to the evolution of
the energy density, and thus to the expansion of the universe. It is seen that the equation-of-state parameter is actually increasing during the
initial stages of evolution. However, at later times, the equation-of-state parameter decreases and becomes negative. Finally, at very late times, it increases and saturates to a negative value. It is observed that $w \ge -1$. This guarantees the stability of the theory and the saturation towards $w \ge -1$ is
\textit{considered to be} resulting from the coupling of Gauss-Bonnet term with the non-Dirac fermionic field.
\subsection{Final Remarks and Conclusions}

In this work we have considered a cosmological model inspired by string theory in which Gauss-Bonnet term is coupled to the non-Dirac fermionic field. Here
the self-interaction potential is considered as a combination of the scalar and pseudo-scalar invariants.

We have investigated the possibility whether the coupling of Gauss-Bonnet term with a non-Dirac fermionic field--characterized by an interaction term $L_{DG}^2$--contributes significantly to the cosmological evolution. We have observed that the new type of coupling plays an important role in the
accelerating behavior (both early and late time) of the universe. It is seen that the behavior of the equation-of-state parameter $\left( w \right)$ is such that $w\ge -1$, which guarantees the stability of the theory.

Hence we conclude that the coupling of Gauss-Bonnet term with the non-Dirac fermionic field has a significant effect on the cosmological
evolution. An interesting feature about this model is that this can generate standard acceleration similar to Dirac as well as a non-Dirac behavior by
simply adjusting the magnitude of the coupling constant $a_2$.

\section*{Acknowledgement}

The authors gratefully acknowledge useful discussions with Urjit A. Yajnik and his keen interest and help in the preparation of this manuscript.

\end{document}